\documentclass[preprint,12pt]{elsarticle}

\newcommand{\specialcell}[2][c]{%
  \begin{tabular}[#1]{@{}c@{}}#2\end{tabular}}

 \usepackage{graphicx}
 
 \usepackage{float}
 
 \usepackage{algorithm}
 \usepackage{algorithmic}
\usepackage{color}
\usepackage{hhline}
\usepackage{mathrsfs}
\usepackage{graphicx}
\usepackage{dcolumn}
\usepackage{bm}% bold math
\usepackage{multirow}
\usepackage{booktabs}
\usepackage{afterpage}
\usepackage{amsmath}
\usepackage{ulem}

\usepackage{tablefootnote}

\usepackage{hyperref}

\newcounter{bla}

\journal{Computer Physics Communications}

\begin{document}

\newpage

\begin{frontmatter}



%\title{A Parallel Algorithm of Nonempirical Hyperdynamics for Seeking the Reaction Minimum-Energy Paths with application for the Argon Clusters}

%\title{AtomREM: A Program Parallel Implementation for Non-empirical Weighted Langevin Mechanics on the Inverse Potential Landscape}
\title{AtomREM: Non-empirical seeker of the minimum energy escape paths on many-dimensional potential landscapes without coarse graining}

\author[a]{Yuri S. Nagornov\corref{author}}
\author[a]{Ryosuke Akashi}
%\author[b]{Third Author}

\cortext[author] {Corresponding author.\\\textit{E-mail address:} iurii@cms.phys.s.u-tokyo.ac.jp}
\address[a]{Department of Physics, The University of Tokyo, Hongo, Bunkyo-ku, Tokyo 113-0033, Japan}
%\address[b]{Second Address}

\begin{abstract}
Recently a non-empirical stochastic walker algorithm has been developed to search for the minimum-energy escape paths (MEP) from the minima of the potential surface [J. Phys. Soc. Jpn. \textbf{87}, 063801 (2018); Physica A, \textbf{528}, 121481 (2019)]. This method is based on the Master equation for the distribution function of the atomic configuration which has a nature to seek the MEP up along the valley of the potential surface. This paper introduces AtomREM (Atomistic Rare Event Manager), which is an MPI parallelized solver program package for executing this method, which yields minimum energy reaction pathways in terms of the microscopic evolution of atomic positions. It is open-source and released under the GNU General Public License (GPL). AtomREM interfaces with the LAMMPS Molecular Dynamics Simulator as a library of versatile potential functions for application to various systems. Examples of the applications to molecular and solid systems are presented.
\end{abstract}

\begin{keyword}
 Potential landscape \sep Reaction paths \sep Minimum-energy escape paths \sep Langevin mechanics \sep Non-empirical scheme  \sep Stochastic algorithm \sep Parallel MPI implementation.
\end{keyword}

\end{frontmatter}

% \linenumbers
%\newpage
% Computer program descriptions should contain the following
% PROGRAM SUMMARY.

{\bf PROGRAM SUMMARY}
  %Delete as appropriate.

\begin{small}
\noindent
{\em Program Title:}   AtomREM (Atomistic Rare-Event Manager)                                         \\
{\em Program Files:}   \href{https://github.com/ryosuke-akashi/AtomREM}{https://github.com/ryosuke-akashi/AtomREM}                                         \\
{\em Licensing provisions: }  GNU General Public License 3 (GPL)                   \\
{\em Programming language:}    Fortran 90 and C                \\
\noindent
{\em Supplementary material:}     1. Program code of AtomREM; 2. Results of simulation for argon solid and $(CH_2)_4$ molecule                                \\
  % Fill in if necessary, otherwise leave out.
%{\em Journal reference of previous version:}                  \\
  %Only required for a New Version summary, otherwise leave out.
%{\em Does the new version supersede the previous version?:}   \\
  %Only required for a New Version summary, otherwise leave out.
%{\em Reasons for the new version:}\\
  %Only required for a New Version summary, otherwise leave out.
%{\em Summary of revisions:}*\\
  %Only required for a New Version summary, otherwise leave out.
\noindent
{\em Nature of problem:} 
AtomREM has been developed to help to find a reaction path on the atomic level using weighted Langevin mechanics on the inverse potential landscape. 
The method is describe the low temperature transformation of complex systems without artificial forces or/and collective variables. 

\noindent
{\em Solution method:}
Recently [1] we designed a non-empirical scheme to search for the minimum-energy escape paths from the minima of the potential surface to unknown saddle points nearby with one dimensional application. The method is based on the Master equation and its solver algorithm is constructed to move the walkers up the surface through the potential valleys [2]. The stochastic algorithm uses a birth/death stochastic processes for numerous of walker in combination with the Langevin equation. Each walker obey the statistics of average movement on the reaction path under biasing potential. Under this consideration the reaction path is derived as the average of the walker distribution from stable to saddle states.

%\\
% {\em Additional comments including Restrictions and Unusual features (approx. 50-250 words):}\\

%* Items marked with an asterisk are only required for new versions
%of programs previously published in the CPC Program Library.\\
\end{small}

\newpage
%\section{}
%\label{}

%{\bf Abbreviation}

%Minimum Energy Path - MEP

%Program Code - AtomREM

%Nonempirical Hyperdynamics - NMD 

%Saddle point - SP 

\section{Introduction}
Searching optimal values of bounded functions defined for many dimensional space concerns diverse scientific problems. In the context of physics (and also chemistry and biology), minimization of the potential energy with respect to the atomic positions is the representative of such problems--the optimal atomic configurations are related to (meta)stable states of molecules and solid systems. A closely related important problem is to locate the paths connecting two (meta)stable states with minimum potential energy barrier, as it specifies the possible chemical reactions and, more generally, structural deformations, as well as its approximate rate of 
occurrence as represented by the Arrhenius equation~\cite{Laidler-text} . A situation where the destinations of the paths are unknown is especially interesting since it concerns seeking of reactions and deformations to form unknown products, which is the main focus of the present article.

Most existing methods to seek such reaction pathways using molecular dynamics and mechanics use collective variables and/or artificial forces; namely, enhance the movement of the atoms in certain targeted directions so that the desired reaction proceeds~\cite{Carter-Ciccotti-bluemoon,Sprik-Ciccotti-bluemoon,Schlitter-Wollmer-targetedMD,SteeredMD-orig-Sci1996,Voter-hyperdyn-JCP1997,Voter-hyperdyn-PRL1997,Laio01102002,Darve-ABF-JCP2001}. However, those approach requires prior knowledge or assumption of the products to designate the collective variables or artificial forces and the solutions are inevitably subject to the human bias.
  
Apart from this mainstream, we have proposed a method based on the Langevin mechanics to seek the paths from known to unknown minima with minimum potential barriers without the collective variables or artificial forces~\cite{Akashi}. This method has been demonstrated to be useful for seeking the reaction paths of the argon clusters with several tens of particles~\cite{Nagornov}, which shows its potential broad applicability to systematic search for unknown reaction pathways described by atomic configurations. The purpose of the present article is to share the current version of the code package for this method as ``AtomREM~\cite{AtomREM}"--Atomistic Rare Event Manager. The current version provides the following functionalities:
\begin{itemize}
\item Nonempirical simulation of the paths from the known to unknown local minima with practical computational time up to at least several tens of atoms
\item Output of the paths visualizable with open-source softwares such as OVITO~\cite{Stukowski_2009,OVITO}.
\item Flat-MPI parallel execution for acceleration of the calculation
\item Use of LAMMPS~\cite{LAMMPS} as an external library for various potential functions.
\end{itemize}

In Sec.~\ref{sec:theory} we briefly explain the background theory and methods implemented in the current version of AtomREM published previously in Refs.~\cite{Akashi, Nagornov}. The contents and usage of AtomREM is described in Secs.~\ref{sec:contents} and \ref{sec:use}, respectively. Some demonstrations are provided in Sec.~\ref{sec:examples}. Section \ref{sec:conclusion} is devoted to concluding remarks. Summary of the source codes and detailed discussion of the numerical behavior are also provided in Appendices for future developers.

\section{Theory and method}
\label{sec:theory}
Suppose any potential surface $U({\bm x})$ is defined on the space of configuration of $N$ atoms [${\bm x}\in {\bf R}^{3N}$; ${\bm x}=(x_{1}, x_{2}, \dots, x_{3N})$]. The task of AtomREM is to generate continuous trajectories non-empirically from a known local minimum of $U({\bm x})$ to unknown minima across the saddle points, on the basis of the overdamped Langevin equation
\begin{eqnarray}
dx_{i} =  -\frac{\partial_{i} U({\bm x})}{\Gamma} dt+ \sqrt{\frac{2k_{\rm B} Tdt}{\Gamma}}W_{i}
,
\label{eq:SDE-x}
\end{eqnarray}
where $x_i$ and $i$ are the coordinate and the index for the degrees of freedom,
${\bm W}=(W_{1}, W_{2}, \dots, W_{3N})$ is the vector whose components are randomly generated from the standard normal distribution at each step, 
$\Gamma$ is the friction constant,
$dt$ is a timestep. At low temperature, the dominant rare trajectories across the saddle points nearby hardly occurs with straightforward simulation of Eq.~(\ref{eq:SDE-x}) because of the exponential dependence of the rate of occurrence $\propto {\rm exp}[-\delta E/T]$ with $\delta E$ being the potential barrier height. On the other hand, under high temperature the atomic configuration is randomly perturbed strongly and information of the trajectory is lost or even system could be melt. AtomREM manages the generation of monotonic trajectories across the saddle points utilizing a recently proposed stochastic process~\cite{Akashi, Nagornov} which has a nature of tracking up the minimum energy paths (MEP) to the saddle points.

\begin{figure}[!t]
 \begin{center}
 \includegraphics[scale=0.36]{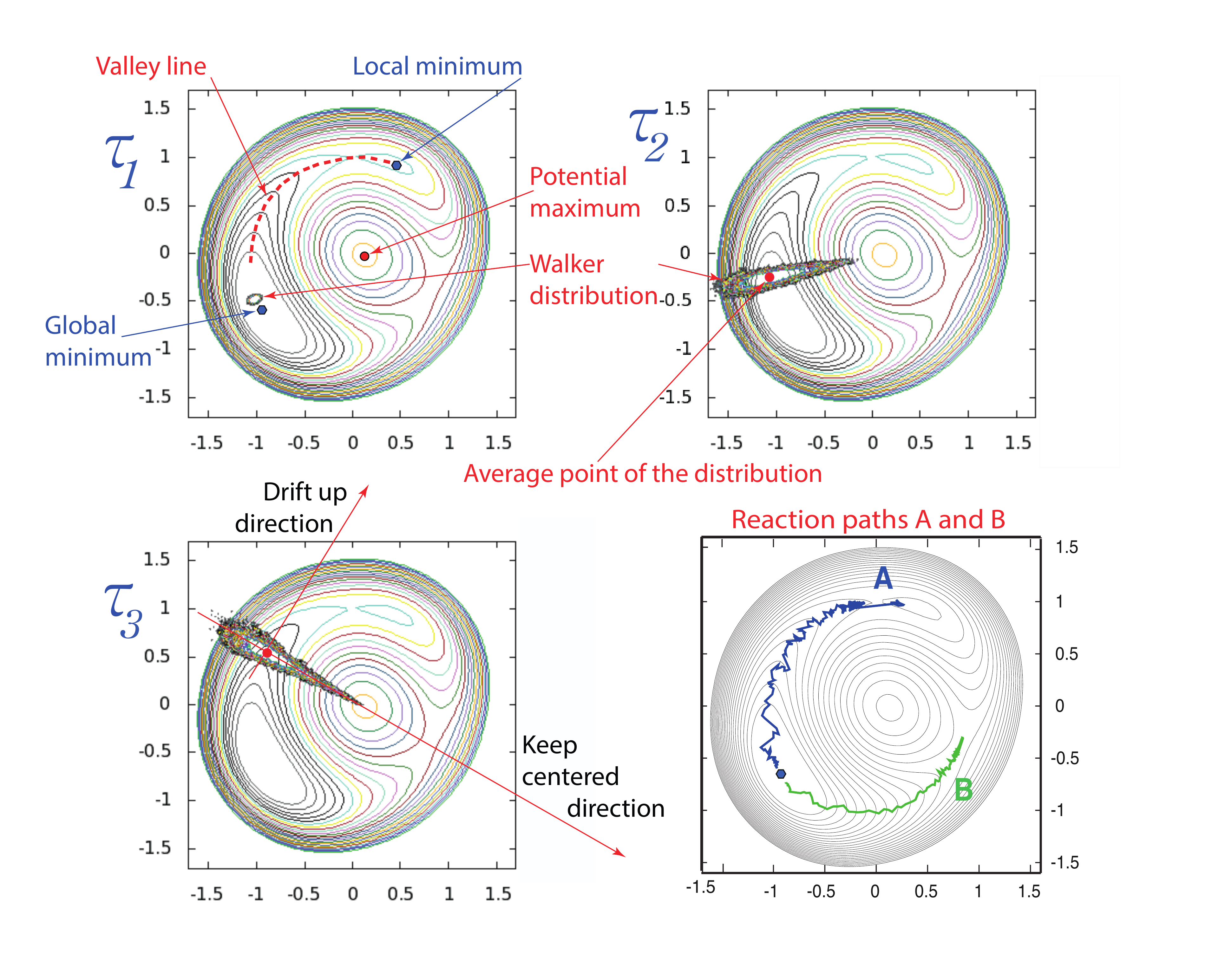}
  \caption{(Color online)   Snapshots of the time evolution of the biased distribution function $q({\bm x},t)$ during the simulation with 2D potential surface $U(x,y)$ 
    in the logarithmic scale at time $t=\tau_1 < \tau_2 < \tau_3 $ \cite{Nagornov}.  The right down snapshot shows the reaction pathways {\bf A} and {\bf B} as the average of the walker positions, which are started from different entrances near the minimum point \cite{Akashi}. }
  \label{fig:valleytrack}
 \end{center}
\end{figure}

Here we briefly explain the background theory on the method. The rarity of the events across the saddle points with Eq.~(\ref{eq:SDE-x}) can also be interpreted with the equivalent Fokker-Planck (Smoluchowski) equation~\cite{Gardiner-book}
\begin{eqnarray}
\partial_t p({\bm x},t) &=& \frac{1}{\Gamma}\partial_i [ (\partial_{i}U({\bm x})) + k_{B}T \partial_i]p({\bm x},t)
\\
&\equiv& \hat{L}_{\rm Sm}p({\bm x},t)
.
\label{eq:Smoluchowski}
\end{eqnarray}
This is the equation that describes the probability distribution $p({\bm x},t)$ of the variable ${\bm x}$ at time $t$ evolved by the Langevin equation Eq.~(\ref{eq:SDE-x}). In this representation, the rarity of the saddle crossing is interpreted as small amplitude of $p({\bm x}, t)$ at the saddle points. The rarity in the long-time simulation is inferred by the Boltzmann distribution $p_{\rm B}({\bm x})\propto {\rm exp}[-U({\bm x})/k_{\rm B}T]$ as the stationary solution of Eq,~(\ref{eq:Smoluchowski}). One can therefore conceive that, with a localized initial condition $p({\bm x}, t_{0})=\delta({\bm x}-{\bm x}_{0})$, the quantity $p({\bm x},t)/p_{\rm B}({\bm x})$ must have sizable components that spread toward high-energy regime.

On the basis of the above consideration, the authors have examined the behavior of the distribution function
\begin{eqnarray}
q({\bm x}, t) \equiv p({\bm x}, t)/{\rm exp}[-V({\bm x})/k_{\rm B}T]
\end{eqnarray}
with $V({\bm x})\equiv (1-\eta)U({\bm x})$ and $\eta$ being the control parameter. In the one dimensional harmonic potential case ($U({\bm x})=U(x)\propto x^2$)~\cite{Akashi}, the following properties have been found with the initial condition $p(x, t_{0})=\delta(x-x_{0})$: 
(i) it has the Gaussian form and its width gradually spreads with t; 
(ii) if $0< \eta < 1/2$, its center goes up the potential surface in a short time and comes back to the potential minimum, whose initial velocity is proportional to $x_{0}$. The authors have demonstrated that these properties can be utilized to track up the ``valley line (Fig.~\ref{fig:valleytrack})" of general potential surfaces from the potential minima~\cite{Akashi, Nagornov}. Namely, in many dimensions, if we set the initial condition $p({\bm x}, t)=\delta({\bm x}-{\bm x}_{0})$ with ${\bm x}_{0}$ in the middle of the valley line, the distribution $q({\bm x}, t)$ goes up along the line keeping the well-localized Gaussian form. One can therefore draw the valley line extending toward the saddle point by tracking the short time evolution of the many-dimensional distribution $q({\bm x}, t)$. The authors have indeed shown the successful visualization of the trajectories connecting the potential minima through the saddle points for argon clusters~\cite{Nagornov}.

In summary, AtomREM solves the Master equation of $q({\bm x}, t)$
\begin{eqnarray}
q({\bm x},t+\tau)
={\rm exp}\{[\hat{L}'_{\rm Sm}+\hat{L}_{\rm rate}] \tau\}q({\bm x},t)
\label{eq:Master-eq}
\end{eqnarray}
with
\begin{eqnarray}
\hat{L}'_{\rm Sm}
&=&
\frac{1}{\Gamma}
\partial_{i} [\partial_i (U({\bm x})-2V({\bm x}))]+\frac{k_{\rm B}T}{\Gamma}\partial_{i}^2,
\\
\hat{L}_{\rm rate}
\equiv
L_{\rm rate}({\bm x}, t)
&=&
\frac{1}{\Gamma}
\left[
F({\bm x})-\langle F\rangle_{q(t)}
\right],
\label{eq:L-rate-def}
\\
\partial_t {\rm ln}C(t) &=& \frac{1}{\Gamma}\langle F\rangle_{q(t)},
\label{eq:C-timeevol}
\\
F({\bm x})
&=&
\partial_{i}^{2}V({\bm x})
+
\frac{1}{k_{\rm B}T}(\partial_{i}V({\bm x}))[\partial_{i}(V({\bm x})-U({\bm x}))]
.
\label{eq:Ffunc}
\end{eqnarray}
Here $C(t)$ is the normalization coefficient for keeping the norm of $q$ unity. The average of the function $f({\bm x})$ is defined as $\langle f\rangle_{q(t)} = \int d{\bm x} q({\bm x},t)f({\bm x})$. The dynamics of $q({\bm x}, t)$ is recast to the time evolution of walkers, which respectively have ${\bm x}$ with different values and their assembly forms $q({\bm x}, t)$. By the Suzuki-Trotter decomposition~\cite{Suzuki-Trotter-1-AMS1959,Suzuki-Trotter-2-CMP1976} Eq.~(\ref{eq:Master-eq}) can be implemented as the repetition of usual Langevin evolution on a modified potential surface and stochastic copying and removal of walkers; the following formula is adopted in AtomREM 
\begin{eqnarray}
&&{\rm exp}\{[\hat{L}'_{\rm Sm}+\hat{L}_{\rm rate}] \tau\}
\nonumber \\
&&\simeq
{\rm exp}\{\hat{L}'_{\rm rate} \tau/2\}
{\rm exp}\{\hat{L}_{\rm Sm} \tau\}
{\rm exp}\{\hat{L}'_{\rm rate} \tau/2\}
+ O(\tau^3)
.
\label{eq:timestep-decompose}
\end{eqnarray}
The operator ${\rm exp}\{\hat{L}_{\rm Sm} \tau\}$ corresponds to the independent time evolution of the walkers by
\begin{eqnarray}
&&{\bm x}(t+\tau)-{\bm x}(t) \nonumber \\
&&=-\frac{1}{\Gamma}\nabla (U({\bm x}(t))-2V({\bm x}(t)))\tau +  \sqrt{\frac{2k_{\rm B}T\tau}{\Gamma}} {\bm W}.
\label{eq:Lang_step}
\end{eqnarray}
Note that, compared with the original Langevin equation Eq.~(\ref{eq:SDE-x}), the potential surface is modified from $U({\bm x})$ to $U({\bm x})-2V({\bm x})$ and, with $0<\eta<1/2$, the modified potential is proportional to the inverse of $U({\bm x})$. This is the main driving force that makes the walkers go up the potential surface. The operator ${\rm exp}\{\hat{L}'_{\rm rate} \tau/2\}$ is a multiplication of a scalar function to $q({\bm x}, t)$ and implemented as the copying and removal of the respective walkers with the probability depending on the magnitude of the scalar function. For this task, an algorithm keeping the number of walkers strictly constant~\cite{conservation_walkers} has been implemented in AtomREM~\cite{Nagornov}. Parallel execution of the whole evolution step [Eq.~(\ref{eq:timestep-decompose})] is available.

We have explained above how the dynamics of $q({\bm x}, t)$ is executed with a {\it given} initial position ${\bm x}_{0}$. Another important task of AtomREM is to generate the initial points that connect to various saddle points with low energy barriers. For this purpose ``heat and relax" method~\cite{Nagornov} is used. Generally speaking, the relaxation of ${\bm x}$ by the steepest descent method behaves like the two-step process; fast relaxation onto the valley line and slow relaxation along it. One can therefore expect that, among randomly distributed (heated) walkers, the walkers which come late across any small energy threshold are expected to be on the valley lines. AtomREM also provides the function to get the initial positions which are likely to connect to low-energy saddle points on this principle. A simple description of the procedure will be appended later, though more detailed information is available in Ref.~\cite{Nagornov}.

%\section{Software implementation}
\section{Contents of the package}
\label{sec:contents}
AtomREM provides three functions:
\begin{enumerate}
\item
{\it Langevin}--solve the overdamped Langevin equation Eq.~\ref{eq:SDE-x}, mainly for the purpose of finding local minima of the given potential function $U({\bm x})$

\item
{\it Initialization}--find the initial positions to start the tracking of the MEP of $U({\bm x})$

\item
{\it Reaction}--track the MEP of $U({\bm x})$ from the given initial positions toward the saddle points

\end{enumerate}
It is composed of two directories, where the source codes, interfaces, and sample inputs for those functions are contained.
\begin{enumerate}

\item
{\it lammps\_pot/}--package implemented for utilizing the potentials of LAMMPS package (recommended)

\item
{\it analytic\_pot/}--package implemented with in-house analytic formula for the potential (only the Lennerd-Jones potential with open boundary condition is available in the current version)

\end{enumerate}

\subsection{Dependencies on external softwares}

For usage of the potential functions implemented in LAMMPS code, AtomREM requires the users to build LAMMPS as a static library (for details see Sec. 3.4 Basic build options in LAMMPS manual page~\cite{LAMMPS}). Also, the external codes summarized in Table: \ref{table:exter} are modified and redistributed in AtomREM. 

\begin{table}[!t]
\caption{\bf External softwares}
\begin{center}
\begin{tabular}{|c|p{10cm}|}
\hline
  {\it mt19937.f90} & Subroutine generating random numbers by the Mersenne Twister method \cite{MersenneTwister}, distributed in {\rm http://www.math.sci.hiroshima-u.ac.jp/\verb|~|m-mat/MT/VERSIONS/FORTRAN/fortran.html}
\\ \hline
 {\it libfwrapper.c} & Wrapper for calling lammps from fortran code. This code is a modification of the one provided in LAMMPS (examples/COUPLE/fortran/) 
 \\ \hline
 {\it USER-LAPLACIAN/} & Package for calculating Laplacian to be installed in LAMMPS. This package is a modification of {\it LAMMPS\_LOCAL\_HESSIAN} package by S. Kadkhodaei~\cite{localHessian}, 
 distributed in {\rm https://cmrl.lab.uic.edu/lammps\_hessian.html}.\\
\hline
\end{tabular}
\end{center}
\label{table:exter}
\end{table}

\section{Usage}
\label{sec:use}
AtomREM has to be built and used in Linux system. This software has been verified to run in CentOS6 systems with Intel compiler. Although in other systems the regular running is not guaranteed as is, we believe that it can be made run regularly by small modification of Makefile.
\subsection{Installation}
Import the source code by {\it git}.\\
{\tt \$ git clone https://github.com/ryosuke-akashi/AtomREM}\\
Enter either of the directories, {\it lammps\_pot} or {\it analytic\_pot}, edit {\it Makefile} for the proper compilers and include and library paths. and enter the following.\\
{\tt \$ make}\\
If the compilation is properly finished the executable {\it a.out} is made.

For {\it lammps\_pot/}, additional procedure is mandatory before building AtomREM to install the user package of calculating Laplacian for Eq.~(\ref{eq:Ffunc}) in LAMMPS.\\
{\tt \$ cp -r USER-LAPLACIAN {\color{blue}\$\{lammps\_source\_directory\}}}\\
{\tt \$ cd {\color{blue}\$\{lammps\_source\_directory\}}}\\
{\tt \$ make yes-user-laplacian}\\
{\tt \$ make mode=lib mpi}\\
After those steps static library {\it liblammps.a} is obtained. The directory \\ {\tt {\color{blue}\$\{lammps\_source\_directory\}}} must be specified in Makefile of AtomREM for linking this static library.

\subsection{Input description}
\begin{figure}[!t]
 \begin{center}
 \includegraphics[scale=0.5]{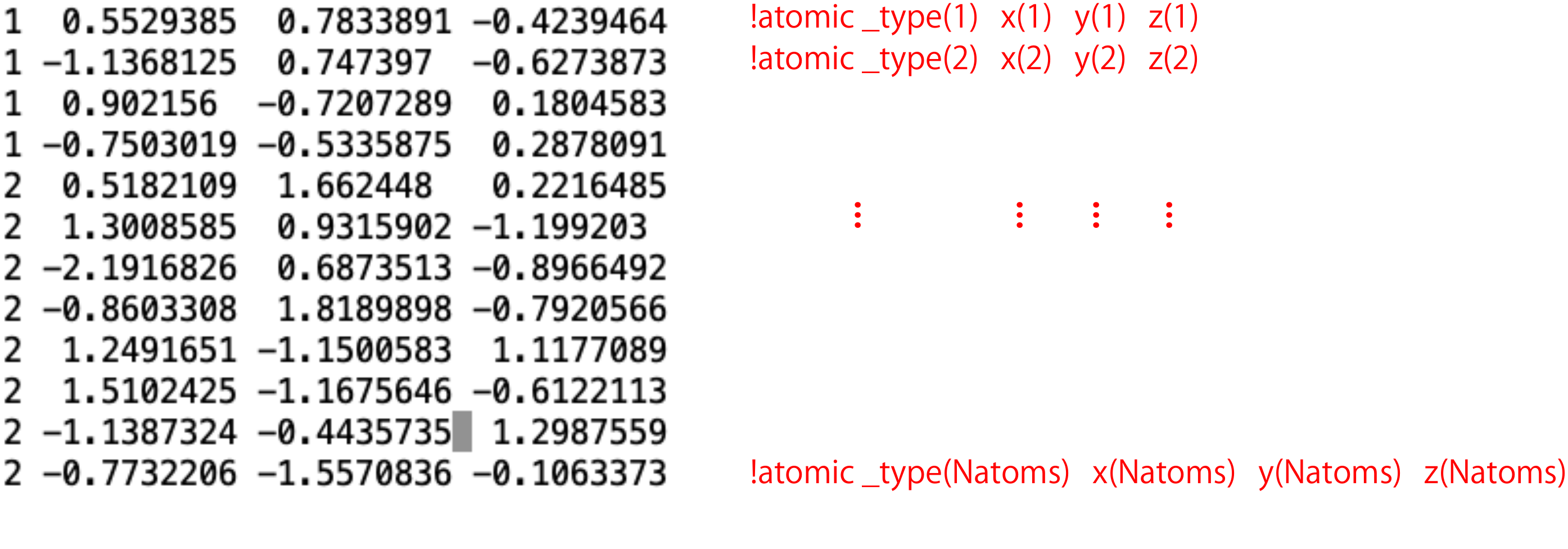}
  \caption{(Color online)  Format of {\it atoms.dat} file defining the initial positions of atoms. An example of the cyclobutane molecule.}
  \label{fig:atomsdat}
 \end{center}
\end{figure}

\begin{figure}[!t]
 \begin{center}
 \includegraphics[scale=0.5]{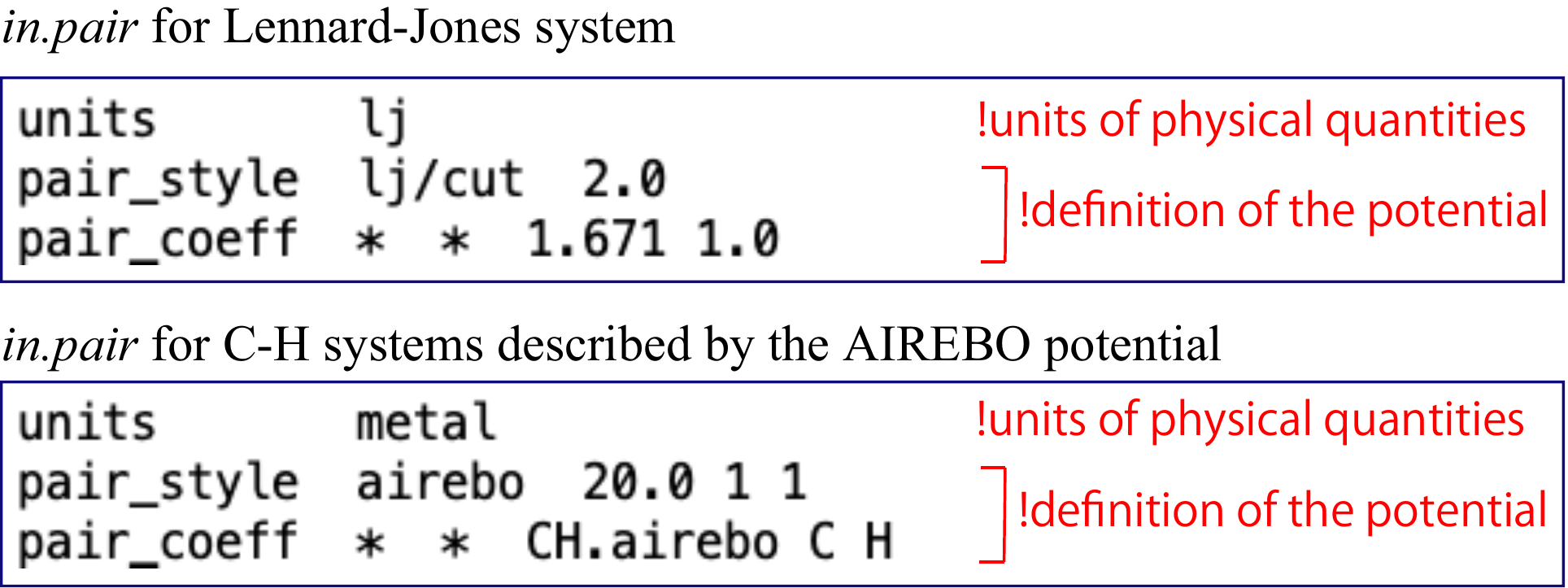}
  \caption{(Color online)  Format of {\it in.pair} file defining the potential form, according to the syntax of LAMMPS commands~\cite{LAMMPS}. Examples of (up) the Lennard-Jones system and (down) hydrocarbon systems described by the AIREBO potential~\cite{airbo}.}
  \label{fig:inpair}
 \end{center}
\end{figure}

AtomREM executables requires three input files:
\begin{enumerate}
\item
{\it params.in}--Input parameters for controlling the calculations written in the fortran namelist format (order insensitive). Imported as a standard input. The parameters are described in Table~\ref{table:parameters}

\item
{\it atoms.dat}--Initial atomic position ${\bm x}_{0}$ to generate the initial distribution $p({\bm x}, t_{0})=\delta({\bm x}-{\bm x}_{0})$. The format is described in Fig.~\ref{fig:atomsdat}.

\item
{\it in.pair}--(Mandatory only for {\it lammps\_pot/}) Part of LAMMPS input file where the potential function is defined. The format is described in Fig.~\ref{fig:inpair}.
\end{enumerate}
The LAMMPS potential file declared in {\it in.pair} must also be put in the working directory. For example, with the lower {\it in.pair} in Fig.~\ref{fig:inpair}, the potential file named ``CH.airebo" is needed.

\subsubsection{Parameters in {\it params.in}}

Most important parameters of the simulation are presented in the file of parameters (Table \ref{table:parameters}), which includes the numbers of walkers and atoms, the time of the simulation by the number of steps and the duration of the time step, temperature and initial value of delta parameter, the names of the input and output files, and also the simulation box. 

The specific parameters for each mode ("Initialization", "Langevin" and "Reaction") will be explained in the corresponding subsections in the section "Using AtomREM". 
\begin{table}[!t]
\caption{\bf Content of parameter file}
\begin{center}
\begin{tabular}{| l | l |}
\hline
{\bf  Example}	&	{\bf Explanation}\\
\hline
\hline
\&input	&	Start line for input file \\
\hline
Nstep = 12000,	&	 Total number of time steps for simulation \\
\hline
Nwalker = 1600,	&	Number of walkers \\
\hline
Natoms = 7,	&	 Number of atoms \\
\hline
temp = 0.1,	&	 \specialcell{ Temperature in the energy units of LAMMPS \\ or analytical potential } \\
\hline
tempFin = 0.4,	&	 \specialcell{ Final temperature in the energy units of LAMMPS \\ or analytical potential } \\
\hline
steptowrite = 50,	&	 \specialcell{ Number of steps to write the atomic coordinates \\ and their energies }\\
\hline
dt  = 0.002,	&	Value of time step  \\
\hline
ratio = 0.48	&	 Value of initial delta function \\
\hline
mode = "Initialization"	&	\specialcell{ Name of mode = "Initialization" \\ or "Langevin" or "Reaction"  } \\
\hline
\hline
& {\bf Next data define the simulation box: } \\
\hline
x\_orig = -8.897999	&	x\_orig,  y\_orig,  y\_orig are the coordinates \\
\hline
y\_orig = -8.3346005	&	 of the down left angle of the box \\
\hline
z\_orig = -6.0167999	&	  \\
\hline
a\_vec1 = 18.5668984 0.0 0.0	&	 a\_vec1, a\_vec2, a\_vec3 are the vectors,  \\
\hline
a\_vec2 = 0.0 17.8101998 0.0	&	which defines the supercell box  \\
\hline
a\_vec3 = 0.0 0.0 11.6429997	&	 \\
\hline
bounds = f f f	&	\specialcell{ Boundaries of simulation box \\ in the term of LAMMPS: f - non periodic, p - periodic }\\
\hline
/	&	 Finish line for input file \\
\hline
\end{tabular}
\end{center}
\label{table:parameters}
\end{table}

\subsubsection{Internal parameters}
There are some tuning parameters in the source code. Although those have been optimized empirically so that the users do not have to modify them, some of them will be referred to in Sec.~\ref{sec:running} for custom usage.

\subsection{Output description}
The program has a lot of output files and the number and kind of these files depend on the simulation mode, but the format of files is similar and they have just three main types. The first one is the sequential list of the time step, the energy, the force, the number of walkers, etc., which is related to the time evolution. The example of output data for the reaction path is in Table \ref{table:path}. The 'E\_atoms.dat' file has the similar format: the 1st column is a time step; the 2nd column is the total energy divided by the number of atoms; the 3rd column is the energy of the 1st atom; the 4th column is of the 2nd, \dots etc. The second main type is the atomic coordinates in the format of LAMMPS trajectories (output files with extension *.lammpstrj), which can be processed by visualization tools such as OVITO~\cite{Stukowski_2009,OVITO}. For example, the 'atomic\_coord\_q.lammpstrj' file has the atomic coordinates for reaction pathway given by the averaging with respect to the walkers. More detailed description of such output files for each mode is presented later in Sec.~\ref{sec:running}. Finally, the code has some miscellaneous output files, such as intermediate ones and default output of LAMMPS. Some of them are also explained later.
%  There is a possibility to reduce the number of columns choosing the specific atoms to save data. The files with names 'E\_atoms\_100XXX.dat' (XXX is the number of thread or processor) have the same format as 'E\_atoms.dat' file, but these files appear during the relaxation after reaching the saddle point.  The relaxation stage in the 'Reaction' mode needs not the averaging procedure, that is why we save relaxation paths independently for each thread.  This procedure allows finding the all possible path from saddle point to the all related stable states.  The files 'coordinates\_rank\_100XXX.lammpstrj' (XXX is the number of thread or processor) content the atomic coordinates of these relaxation paths.
%The third type is the files related to the LAMMPS library, like input file for LAMMPS and output files: "data.case", "in.case" and "log.lammps". For the second and third types of file, please see \cite{LAMMPS} website.  The third type is related to the running of LAMMPS and a user need not to analyze the data of these files, that is needed only for the debugging process.  For example, the log file of LAMMPS has information about conflict with input data or an absence of some files or incorrect data in the file, and etc. The 'saddle.dat' file has information about the atomic coordinates at the saddle point, which appears after seeking of the reaction path and before the relaxation stage in the 'Reaction' mode. This file has a simple format: x,y and z coordinates for 1st atom, the next line has information for the 2nd atom, the third line is for the 3rd atom, and etc. 

\begin{table}[!b]
\caption{\bf Format of output file "path.dat"}
\begin{center}
\begin{tabular}{| c | c | c | c | c |}
\hline
\specialcell{ Time \\ step  \\ $\tau$}	&  \specialcell{ Average energy \\ of atoms $E_q$ \\ (from $q$)} &  \specialcell{  Average energy \\ of atoms $E_p$ \\ (from $p$)} & \specialcell{ Force calculated for \\ the average position of \\ the walkers $F_{\rm ave}$ }	 &  \specialcell{ Number  \\ of walkers \\ $N_w$} \\
\hline
\hline
     1  & -0.39392953E+0001  & -0.39392953E+0001  &  0.37627504E+0000   &   1600 \\
\hline
    50  & -0.39394117E+0001  & -0.39394117E+0001   & 0.34776722E+0000 &      1600 \\
\hline
   100 &  -0.39393155E+0001  & -0.39393155E+0001   & 0.44461967E+0000   &   1600 \\
\hline   
   150  & -0.39389018E+0001  & -0.39389018E+0001   & 0.64310856E+0000   &   1600 \\
\hline   
   200  & -0.39386394E+0001 &  -0.39386394E+0001   & 0.73788821E+0000     & 1600 \\
\hline
\hline
\end{tabular}
\end{center}
\label{table:path}
\end{table}

%\section{Downloading and installing of the AtomREM}
%The program code is available at the repository \href{https://github.com/ryosuke-akashi/AtomREM}{AtomREM} \cite{AtomREM}.  It is open-source and released under the GNU General Public License.  The source code is released as a public open software repository that includes detailed documentation and tutorial cases.  The code utilizes the LAMMPS package as a library \cite{LAMMPS} to calculate the potential function and its first and second derivations.  Another alternative is to use an analytical form of the potential function instead of LAMMPS potential to keep a computational cost of simulations. That is why the compilation has two scenarios. The first one is the compilation without the LAMMPS, but using 'potentials.f90' instead 'potentials\_lmp.f90'. The second one is to use LAMMPS as a library and it needs several steps: 

%- to install the LAMMPS;  - to compile the LAMMPS as an MPI executable file with the library mode, that is possible by the command 'make mode=lib mpi' to get the files 'liblammps\_mpi.a' and 'liblammps.a'; - to write the path to the LAMMPS source file directory into the Makefile as a path to the library; - and finally, to make compilation using Makefile.

\subsection{Running AtomREM}
\label{sec:running}
The {\it a.out} file is executed by entering the following line\\
{\tt \$ ./a.out < param.in > out}\\
Specifying {\it param.in} as the standard input is mandatory, whereas the standard output file name is arbitrary. When the calculation regularly ends, the following line is written at the end of the standard output: \\
{\tt End of simulation.}\\
Flat-MPI parallelization is also possible; indeed, the default Makefile is compatible with the intel MPI. A standard parallel execution using {\it mpirun} is, e.g., as follows:\\
{\tt \$ mpirun -np 16 ./a.out < param.in > out}\\

The situation where AtomREM is useful is that the users know a (an approximate) metastable structure realized with a given potential function $U({\bm x})$ and want to seek for unknown transition pathways beyond the potential barriers. To achieve this goal, the users can run AtomREM in the three modes: "Langevin", "Initialization" and "Reaction". The "Langevin" mode is used to locate an accurate local minimum from a given initial position. The "Initialization" mode is used for finding the initial atomic configurations as the starting points to the reaction paths, which are used as the input of the later execution with the ``Reaction" mode. The algorithm of the finding of the initial atomic configurations is described in work \cite{Nagornov}. The "Reaction" mode is used for seeking of the reaction path from the initial atomic configuration. In order to choose the mode, you have to write the name of a mode in the file of the parameters. Below, we append a detailed description of the three steps.

  \subsubsection{Mode "Langevin"}
The usual Langevin equation Eq.~(\ref{eq:SDE-x}) is executed independently for each process. The minimum number of walkers equal the number of the processes. This mode is provided for finding any potential minima closely related to a given atomic configuration. By executing this with low temperature the system reaches to the minimum through the steepest descent path. Although the users of AtomREM is assumed to know at least one metastable configurations of the system before the use of the code, the precise location of the minimum, which is essential for the later modes, could depend on the potential function used. The purpose of this mode is to correct such dependence. The output files of this mode are only  ’E\_atoms\_100XXX.dat’ and ’coordinates\_rank\_100XXX.lammpstrj’, where 'XXX' is the index of the processes. This mode is also internally called in mode ``Reaction" after reaching the saddle point, to let the walkers move to other metastable configurations beyond the saddle point.
% That allows to see the metastable state as a final point of the simulation, and if the simulation is correct, then the all trajectories go to the same state.

  \subsubsection{Mode "Initialization"}
\begin{figure}[!t]
 \begin{center}
 \includegraphics[scale=0.75]{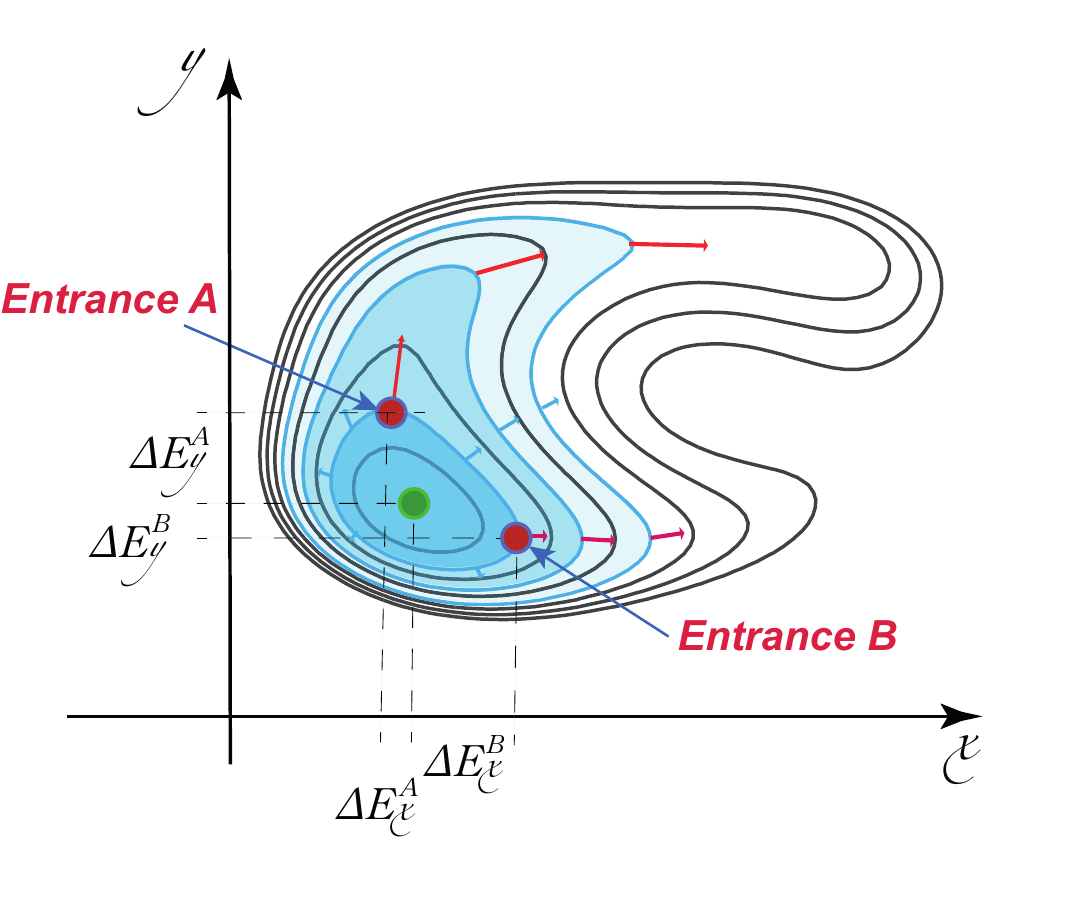}
  \caption{(Color online) 2D example of the potential surface with two entrances to the reaction paths: A and B. There are the conditions to distinguish the entrances: $\Delta E_x^A~<~\Delta E_x^B$ and $\Delta E_y^A~>~\Delta E_y^B $ }
  \label{fig:entrances}
 \end{center}
\end{figure}

%In Appendix B there is the explanation of how the energy deviations of atoms launch the reaction.  
Using the ``heat and relax" algorithm described in the paper \cite{Nagornov}, the mode "Initialization" generates initial atomic configurations that are supposed to be well on the valley lines toward the saddle points, or, ``entrances" to the reaction paths [Fig.\ref{fig:entrances}]: Entrance A is a state with the largest energy deviation to the $y$-direction, and entrance B is a state with the largest energy deviation to the $x$-direction. Note that one initial configuration in principle corresponds to one reaction path.%, and this is a criterion on how to choose the parameters of the initialization. 

\begin{figure}[!t]
 \begin{center}
 \includegraphics[scale=0.75]{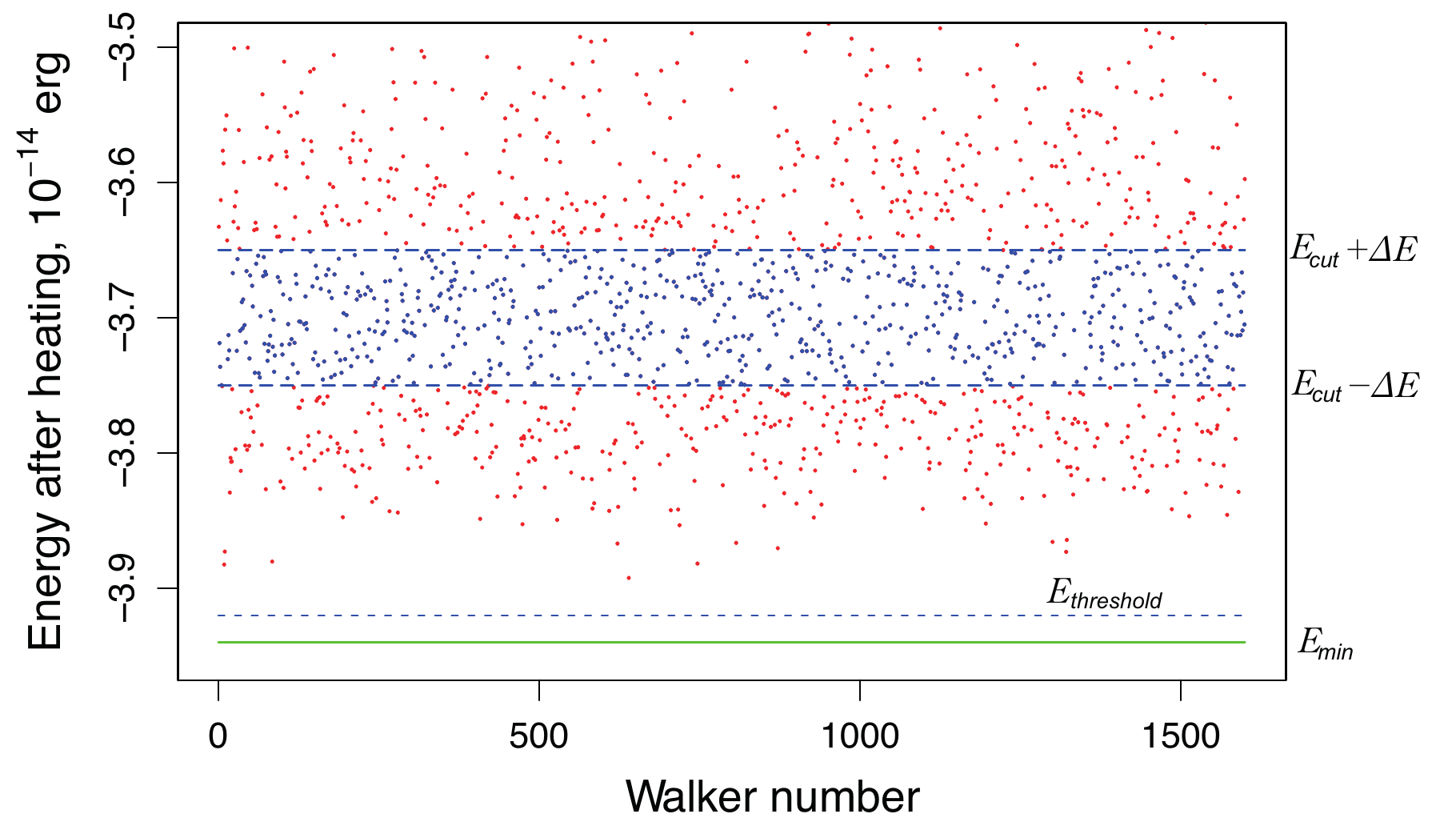}
  \caption{(Color online) Example of energy distribution for $LJ_7$ cluster after heating. 
  The $E\_cut=-3.7$ is a cutoff energy to select the walkers in the range $[E_{cut}~-~\Delta E~,~E_{cut}~+~\Delta E]$ ($\Delta E = 0.05$) after heating of the walkers. 
The $E_{threshold}=-3.92$ is an energy closed to the stable state with energy minimum $E_{min}=-3.94$.}
  \label{fig:Ecut}
 \end{center}
\end{figure}

The task with this mode is to evolve the {\it Nwalkers} walkers with the Langevin equation Eq.~(\ref{eq:SDE-x}) first under temperature {\it temp} for duration {\it Nstep} and later under temperature {\it tempFin} ($\ll${\it temp}) to relax the walkers along their steepest descent paths. When the temperature is switched, the potential energies of the respective walkers $E_{i} (i=1, 2, \dots, Nwalkers)$ are checked. Among the walkers, only those whose energies satisfy the criterion $E_{\rm cut}-\Delta E < E_{i} < E_{\rm cut}+\Delta E$ are kept and others are killed~(Fig.~\ref{fig:Ecut}). Afterwards, the time steps are measured for the remaining walkers to come down to an energy threshold $E_{\rm threshold} (< E_{\rm cut}-\Delta E)$, so that the slowly relaxing walkers are detected. The measured time steps and atomic configurations of the walkers are recorded when they reach to $E_{\rm threshold}$ for later data processing. The parameters for the criteria are set internally, according to the initial temperature {\it temp} in {\it main.f90}.\footnote{The default values are $E_{\rm cut}=E_{\rm ini}+1.5 \times temp$, 
$\Delta E=0.25\times temp$, 
$E_{\rm threshold}=E_{\rm ini}+0.15\times temp$, 
with $E_{\rm ini}$ being the potential energy at the start of this mode. 
One is recommended to modify the definition of those parameters with observation of 
{\it After\_heating\_energy.dat} 
so that many walkers are within the range 
[$E_{\rm cut}-\Delta E, E_{\rm cut}+\Delta E$]}

The mode "Initialization" yields several output files to help the choice of initial atomic configurations. See Table \ref{table:ini} for details. The main information in the output files are (i) the indices of the walkers kept at the temperature switching, (ii) the numbers of time steps required for passing through $E_{\rm threshold}$, (iii) the atomic configurations at the step of reaching $E_{\rm threshold}$, and (iv) the indices of atoms whose per-atom energies (explained later) show the largest deviations from their initial values. The last information can be utilized to a clustering analysis for obtaining the entrance configurations to diverse reaction paths~\cite{Nagornov}. Generally, the classical potential function can be represented as the sum of two-body, three-body and more-body terms as
\begin{eqnarray}
U({\bm x})
=\sum_{a<b}U({\bm x}^{(a)}, {\bm x}^{(b)})+\sum_{a<b<c}U({\bm x}^{(a)}, {\bm x}^{(b)}, {\bm x}^{(c)})+\dots,
\end{eqnarray}
where ${\bm x}^{(a)}$ is the position of the $a$th atom. This form allows one to define the per-atom energy of the $a$th atom by
\begin{eqnarray}
E^{(a)}
\equiv
\sum_{b}U({\bm x}^{(a)}, {\bm x}^{(b)})+\sum_{b<c}U({\bm x}^{(a)}, {\bm x}^{(b)}, {\bm x}^{(c)})+\dots \ \  (b, c, \dots \neq a )
\end{eqnarray} 
AtomREM records the atomic indices which show the three largest values of $E^{(a)}$ for each walker and attributes it to a unique number $Un_{i}$ (e.g., if the third, second, and fourth atoms show the largest deviation of $E^{(a)}$ in this order, $Un_{i}=30204$). Choosing the most slowly relaxing walkers for each group of walkers having common $Un_{i}$ can yield the entrance points to the different reaction paths.\footnote{Note that the proper grouping according to $Un_{i}$ generally requires consideration of the spatial symmetry of the system, though it works to some extent without such consideration.} For flexibility we do not provide specific tools for such data analysis, though the atomic configuration of the most slowly relaxing walker of all is written out in file {\it Chosen\_BEST\_walk.dat} for convenience.

%To gain the diversity of the MEPs generated in the later time evolution step, we need to consider the different speeds of relaxations along various paths. For this purpose, we define atomic energy $E_{a} (a=1, 2, \dots, N_{\rm atom})$ with the interatomic potential $U_{ab}$ by $E_{a}=\sum_{b}U_{ab}$ and characterize {\it the diversity} of the valley lines by the set of the atoms showing largest energy deviations $\Delta E_{a}\equiv E_{a} - E_{a}^{\rm min}$ ($E_{a}^{\rm min}$: value of $E_{a}$ with the optimized atomic configuration at the stable state).  Therefore we group the walkers according to the atoms showing largest $\Delta E_{a}$ and find the most slowly relaxing walkers among those groups (Table~\ref{table:ini}). For example, for $LJ_7$ cluster if we have the $\Delta E_1 = 0.01, \Delta E_2 = 0.13, \Delta E_3 = 0.15, \Delta E_4 = 0.1, \Delta E_5 = 0.01, \Delta E_6 = 0.02, \Delta E_7 = 0.02$, then the sorted list of the energy deviations includes the three largest values as $\Delta E_3, \Delta E_2, \Delta E_4$, so the group name will be "324" in accordance with the order of atoms. 
 
\begin{table}[!t]
\begin{minipage}{1.0\linewidth}
\caption{\bf  Output files for mode "Initialization"}
\begin{center}
\begin{tabular}{| c | c | c |}
\hline
\specialcell{ \bf  File name }	&  \specialcell{\bf Columns \\ \bf in the file } &  \specialcell{\bf Description }  \\
\hline
\hline
After\_heating\_energy.dat &	$i_w\footnote{$i_w$ is number of walker}~~~~~E_i$\footnote{$E_i$ is an average energy of walker} 	&	\specialcell{Average atomic energy \\ for each walker} \\
	\hline
Number\_struct.dat	& $i_w$~~~$Un_{i}$\footnote{$Un_{i}$~is a unique number of group}~~~$\tau_{rlx}$\footnote{$\tau_{rlx}$~is a time of relaxation}	&	\specialcell{ Information about \\ groups and relaxation time } \\
	\hline
Structures.lammpstrj	& -	&	\specialcell{ Configuration data \\ in the format of \\ LAMMPS trajectory }  \\
	\hline
Energies\_of\_structures.dat	&  $i_w$ $E_i$ $E_i^{(1)}$ $E_i^{(2)}$ ... $E_i^{(N)}$ \footnote{$E_i^{(1)}$, $E_i^{(2)}$, ... , $E_i^{(N)}$ are the atomic energies for 1-st, 2-nd, ..., N-th atom} 	&	 \specialcell{ Atomic energies and \\ debagging information about all \\ selected walkers } \\
%	\hline
%ALL\_STRUCTURES.dat	&	&	 \specialcell{  } \\
	\hline
\specialcell{ Chosen\_walk\_Structures\\.lammpstrj } 	&	- &	 \specialcell{ Atomic configurations \\ for chosen walkers  } \\
	\hline
Chosen\_walk\_Number	& $i_w$~~~$Un_{i}$~~~$\tau_{rlx}$	&	 \specialcell{ Information about \\ groups and relaxation time \\ for chosen walkers  } \\
	\hline
Chosen\_walk\_Energy.dat	& $i_w$~~$E_i$~~$E_i^{(1)}$~~$E_i^{(2)}$~...~$E_i^{(N)}$	&	 \specialcell{ Energies of each atom \\ for chosen walkers } \\
	\hline
Chosen\_BEST\_walk.dat	 &  -  & 	 \specialcell{ All information about \\ the slowest walker: number of walker \\ and energies of each atom, \\ number of group, time of relaxation, \\ and coordinates of each atom. \\ Last lines are utilizatble as \\atoms.dat for mode ``Reaction". } \\
\hline
\end{tabular}
\end{center}
\label{table:ini}
\end{minipage}
\end{table}

  \subsubsection{Mode "Reaction"}
\label{sec:reaction}
In the mode "Reaction" the walker distribution evolves according to the equation [Eq.~(\ref{eq:Master-eq})] under temperature {\it temp} in order to seek the reaction path to the saddle point from a given initial point. During the simulation, the {\it Nwalker} walkers are equally distributed to the respective processes. After the evolution of timestep {\it Nstep} is finished, all the walkers are gathered to their average position, the number of walkers per process is changed to unity, and afterwards the final relaxation with Eq.~(\ref{eq:SDE-x}) is executed under temperature {\it tempfin} independently for each process. As a result, with {\it Nstep} being large enough for reaching near the saddle point, some walkers go beyond the saddle point and reach to other metastable points. We thus have the reaction path in the form of the time evolution of the walker average of the atomic positions. The related output files are described in Table~\ref{table:reaction}. For example, plotting {\it path.dat} and several {\it E\_atoms\_100XXX.dat}s at the same time yield the figure like Fig.~\ref{fig:energy_argon} given later.%\footnote{The file "atomic\_coord\_q.lammpstrj" in Table~\ref{table:reaction}} to the saddle point\footnote{The file "saddle.dat" in Table~\ref{table:reaction}}.  The time evolution of energy of these configurations is saved into the file "path.dat" (Tables~\ref{table:path} and \ref{table:reaction}).  

\begin{table}[!t]
\begin{minipage}{1.0\linewidth}
\caption{\bf Output files for mode "Reaction"}
\begin{center}
\begin{tabular}{| c | c | c |}
\hline
\specialcell{ \bf File name }	&  \specialcell{\bf Columns \\ \bf in the file } &  \specialcell{\bf Description }  \\
\hline
\hline
atomic\_coord\_q.lammpstrj	& -	&	\specialcell{ Configuration data \\ in the format \\ of LAMMPS trajectory  \\ under biasing potential $V({\bm x})$ } \\
	\hline
path.dat\footnote{please, see Table \ref{table:path}}	& $\tau$~~$E_{q}$~~$E_{p}$~~$F_{min}$ 	&	\specialcell{ Time evolution of \\ the energy and the minimum force \\ of the walker ensemble \\ under biasing potential $V({\bm x})$} \\
	\hline
E\_atoms.dat	& $\tau$~$E_q$~$E_\tau^1$~$E_\tau^2$...$E_\tau^N$	&	\specialcell{ Energies of each atom \\ of the average position \\ at time step $\tau$ \\ under biasing potential $V({\bm x})$ } \\
	\hline
saddle.dat 	& $type_i$~$x_i$~$y_i$~$z_i$\footnote{$type_i$ is a type of $i$-th atom, the $x_i$,~$y_i$,~$z_i$ are coordinates of $i$-th atom.}	&	\specialcell{ Atomic configuration of  \\ a saddle point \\ before the relaxation stage } \\
	\hline
E\_atoms\_100XXX.dat	& $\tau$~$E_q$~$E_\tau^1$~$E_\tau^2$...$E_\tau^N$	&	\specialcell{ Energies of each atom \\ for the XXX thread (walker) \\ at time step $\tau$ \\  at the relaxation stage} \\
	\hline
 \specialcell{coordinates\_rank\_100XXX. \\ lammpstrj} 	&	-	&	\specialcell{ Atomic configurations \\ for the XXX thread (walker) \\  at the relaxation stage } \\
	\hline
\end{tabular}
\end{center}
\label{table:reaction}
\end{minipage}
\end{table}

During the first part of the simulation, the ``reset and pullback" operation~\cite{Nagornov} is automatically executed with an interval defined internally. Specifically, the continuous evolution with Eq.~(\ref{eq:Master-eq}) yields gradual spreading of $q({\bm x}, t)$ and the simulation eventually becomes subject to incorrect departure from the valley line. To mitigate this instability, the distribution is occasionally reset to the delta function and afterwards evolved with the biasing potential $V({\bm x})=0$ (= usual Langevin equation) in a short time to let the walkers back onto the valley line. 

The parameter $\eta$ defining $V({\bm x})$ has a crucial effect on the stability of the simulation~\cite{Akashi, Nagornov}. The necessary condition for tracking up the potential slope is $\eta<1/2$. With too large $\eta$, the walkers quickly stop climbing up the potential slope and go down. Smaller $\eta$ drives the walkers faster and longer up the potential slope, whereas it tends to induce the numerical instability more frequently. To our empirical knowledge the best way is to start with $\eta$ as close to $1/2$ as possible (set by {\it ratio} in {\it param.in}) and reduce it if the walkers do not climb up the potential surface. The following internal algorithm is adopted: (i) Save the energy of the initial configuration $E_{ini}$ and start the simulation with $\eta$ given in {\it param.in}; (ii) make the simulation for $N_{\rm check}$ time steps\footnote{By default we have set $N_{\rm check}=400$}; (iii) check the walker average of the potential energy $E_{q}$; if $E_{q} > E_{ini}$ continue the simulation; else, decrease  $\eta$ to amplify $V({\bm x})$ and start over the simulation again with the new $\eta$.

\newpage

\section{Examples of simulations and validation}
\label{sec:examples}

We have published the examples of the application of a previous private version of AtomREM to the surface reactions of Ar$_7$, Ar$_{13}$, and Ar$_{38}$ in the paper \cite{Nagornov}. Here we will show other types of transitions in the cyclobutane molecule and argon crystal. The example includes incorrect destruction of the molecule, the C-C bond-breaking transformation and collective sliding transitions as a elementary process of the solid-solid transition. The final example especially represents the advantage of the current method that generates the collective transformation without a priori assumption to the reaction coordinate as it is very difficult for solids to choose the collective variable \cite{PhysRevLett.119.015701}.
%We think that is a very interesting example of simulation under too strong biasing potential $V(\bf x)$, which leads to the destroying of the molecule. This example allows demonstrating the significance of delta parameter and initial atomic configuration for reaction path and reaching the saddle point of a high order. The second example demonstrates the path for the cyclobutane molecule with a reaction for $C-C$ bond and relaxation to the four metastable states. And finally, there is an example of solid-solid transition for argon crystal, that is a main advantage of the method because for solid it is very difficult to choose the collective variable \cite{PhysRevLett.119.015701}.

% \subsection{The destroying the cyclobutane molecule under the strong $V(\bf x)$}

\subsection{Cyclobutane}
\begin{figure}[H]
\begin{center}
\begin{minipage}[h]{0.75\linewidth}
\center{\includegraphics[width=1.0\linewidth]{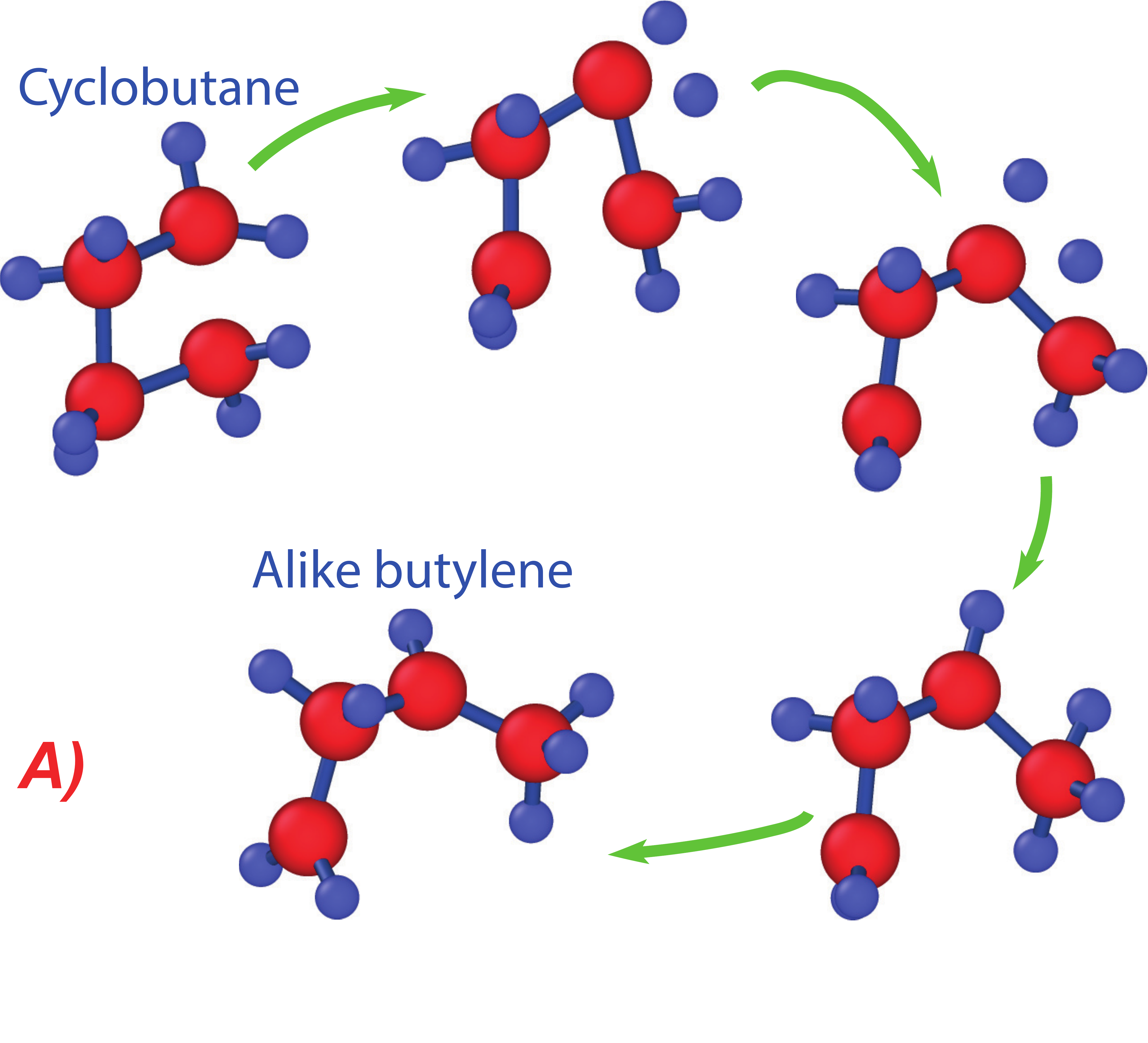}}
\end{minipage}
\hfill
\begin{minipage}[h]{0.75\linewidth}
\center{\includegraphics[width=1.0\linewidth]{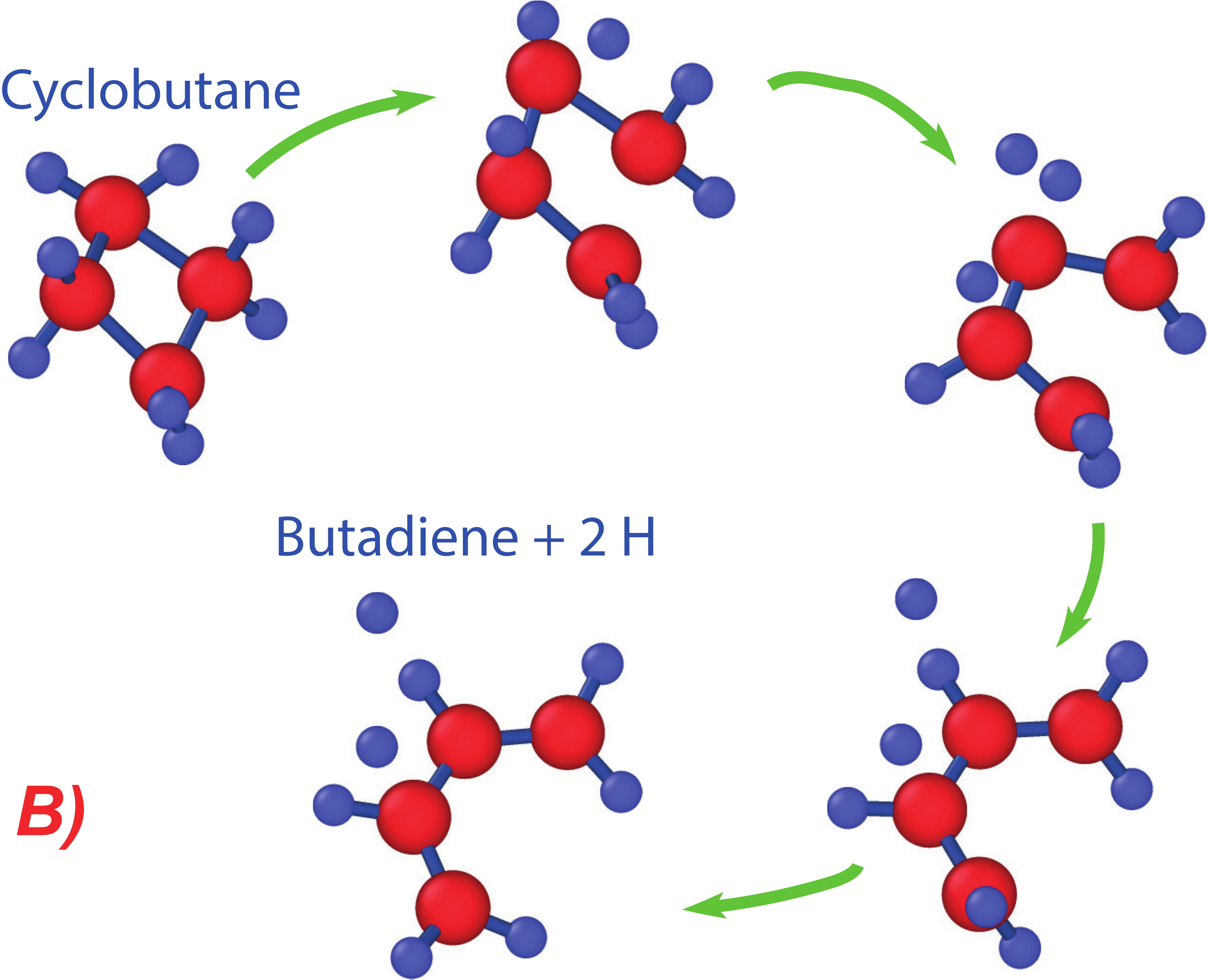}}
\end{minipage}
\hfill 
  \caption{(Color online) The evolution of the cyclobutane structure during the $C-C$ bond breaking:  
  A) and B) are the reaction paths corresponding to the lines 2 and 4 in Fig.~\ref{fig:E_cyclo}.}
  \label{fig:paths_cyclo}
  \end{center}
\end{figure}

We generated the reaction paths of cyclobutane: to butylene-like chain (A in Fig.~\ref{fig:paths_cyclo}), butadiene (B in Fig.~\ref{fig:paths_cyclo}) and tetramethylene (not shown). %The parameters were set as follows: XX, . . . AIREBO, . . . 

\begin{figure}[!t]
 \begin{center}
 \includegraphics[scale=0.75]{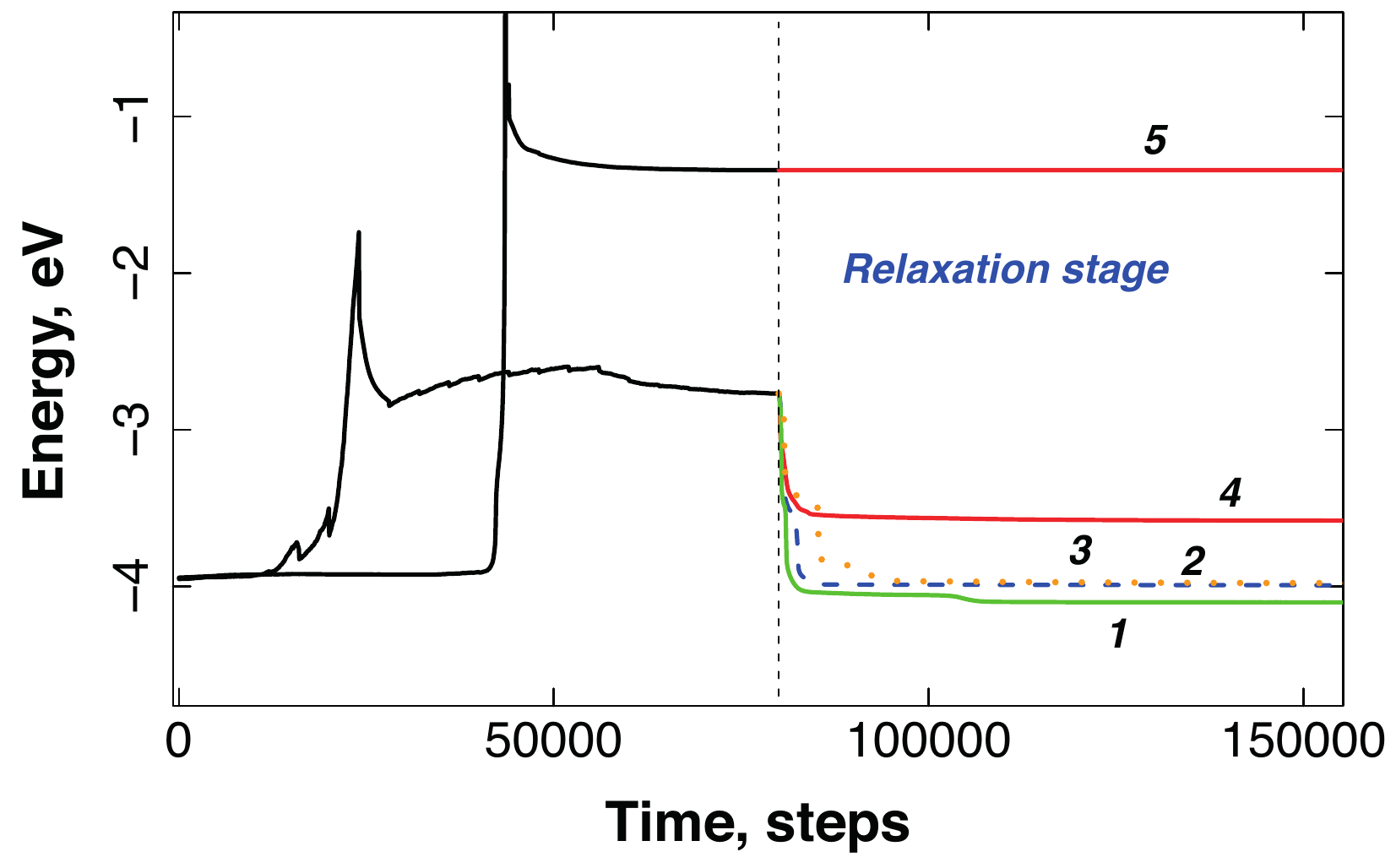}
  \caption{(Color online) The evolution of the energy per atom of cyclobutane for 80~000 time steps under biasing potential $V({\bm x})$ and then the relaxation stage: 
 1-4 lines have the same saddle point of $C-C$ bond breaking, 
 5 line is the simulation with a destroying the molecule structure. 
  The line 1 is the way back to the cyclobutane state, the lines 2 and 4 correspond to the paths A and B in Fig.~\ref{fig:paths_cyclo}., the line 3 corresponds to the tetramethylene formation path, the line 5 corresponds to the path in Fig.~\ref{fig:paths_cyclo_destr}.}
  \label{fig:E_cyclo}
 \end{center}
\end{figure}

In Fig.~\ref{fig:paths_cyclo} the first two snapshots for the A and B paths are the same, where the second is the structure of the saddle point. After reaching the saddle point, the simulation of each process was independently executed under very low temperature ({\it tempfin}=$10^{-4}$ eV) and yielded different paths through the steepest descent paths or Intrinsic Reaction Coordinate (IRC). The only 4 IRC were unique and have led to the initial cyclobutane state (line 1 in Fig.~\ref{fig:E_cyclo}), to the butylene-like chain (A in Fig.~\ref{fig:paths_cyclo}), to butadiene (B in Fig.~\ref{fig:paths_cyclo}) and to tetramethylene (not shown).

The time evolution of the potential energies along these paths extracted from ``path.dat" are shown in Fig.~\ref{fig:E_cyclo}. During the reaction simulation, the lines are serrated due to ``reset and pullback" steps (See Sec.~\ref{sec:reaction}). The reaction pathways are also in the Supplementary movies (see Supplementary materials). 

\begin{figure}[!t]
 \begin{center}
 \includegraphics[scale=0.25]{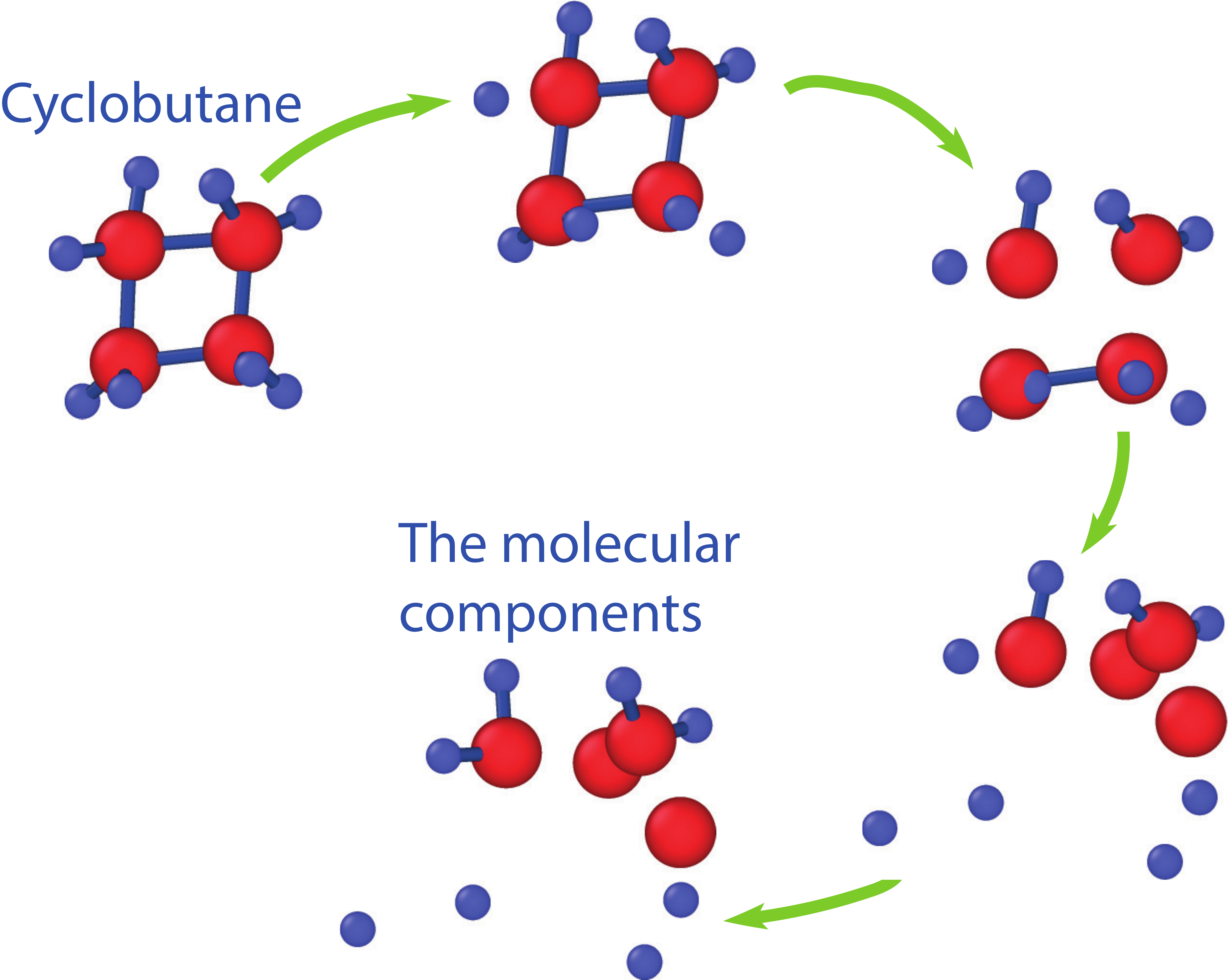}
  \caption{(Color online) The evolution of the cyclobutane structure during the destroying of the molecule. 
  The reaction path corresponds to the line 5 in Fig.~\ref{fig:E_cyclo}.}
  \label{fig:paths_cyclo_destr}
 \end{center}
\end{figure}

If the initial positions are chosen incorrectly, the destruction of the molecule occurs during the simulation (Fig.~\ref{fig:paths_cyclo_destr}) and energy jumps to the high value (line 5 in Fig.~\ref{fig:E_cyclo}). Even if the initial position is correctly on the reaction path, incorrect departure from the path driven by the random fluctuation can occur when the temperature ({\it temp}) is huge. The peak at the 24000-time step for the lines 1-4 (Fig.~\ref{fig:E_cyclo}) is thought to correspond to the latter case; thanks to the reset and pullback step the walkers come back to the proper reaction path.  

Note that the target of the present demonstration is just to show the capability of the method to track the reaction path on a given potential surface, regardless of the accuracy of the potential surface itself. There are a lot of work devoted to the research of cyclobutane, for example \cite{cyclo_laser_2,cyclobutane,cyclo_laser,cyclo_tethra}. In work \cite{cyclo_laser_2} the energy of dissociation was calculated by the composite method CBS-QB3 of quantum chemistry, and intrinsic reaction coordinate calculations have been performed at the B3LYP/6-31G(d) level to ensure that the computed transition states connect the desired reactants and products, whereas we have used the classical potential function AIREBO \cite{airbo}. The energies of the metastable states consequently differ. Also, the positions of the two released hydrogen atoms in the butadiene case is thought to be an artifact of the AIREBO potential, though it could be associated to possible dissociation of the hydrogen molecule H$_2$. %, which has an accuracy of 5-7\% for the energy calculation of the metastable states. For transition state, we calculated the differences between the energy of cyclobutane and the saddle point/peak point are equal ..?.. and ..?..eV respectively.

Fig.~\ref{fig:paths_cyclo_destr} demonstrates the incorrect reaction path with interesting features. The third snapshot shows some hydrogen atoms, which are too close to the carbon atoms because these H atoms drift up their local potential gradients. Due to the too short H-C distance, the energy of the molecule becomes very high (line 5 in Fig.~\ref{fig:E_cyclo}). There are several possible reasons for this behavior: the first is incorrect initial atomic configuration, which does not lie on the reaction path, the second is the high temperature of simulation or high time step, which leads to the high fluctuations, the third is not enough number of walker, which makes the simulation subject to fluctuations of the average position of the walkers. Although the simulation finally settles to fragments of molecules (final snapshot) after the relaxation, the metastability of this specific atomic configuration is obviously the artifact of the model classical potential. This is a typical behavior of the simulation when it fails.

% The reactions on cyclobutane are in the paper \cite{cyclobutane}, 

% the reaction of photodissociation of cyclobutane under laser ablation \cite{cyclo_laser}, \cite{cyclo_tethra}. 

% The reaction path \cite{cyclo_path} by Metadynamics

% The potential AIRBO from paper \cite{airbo} and using the LAMMPS library \cite{LAMMPS}. 

\subsection{The HCP-FCC transition of argon crystal}

\begin{figure}[!t]
 \begin{center}
 \includegraphics[scale=0.7]{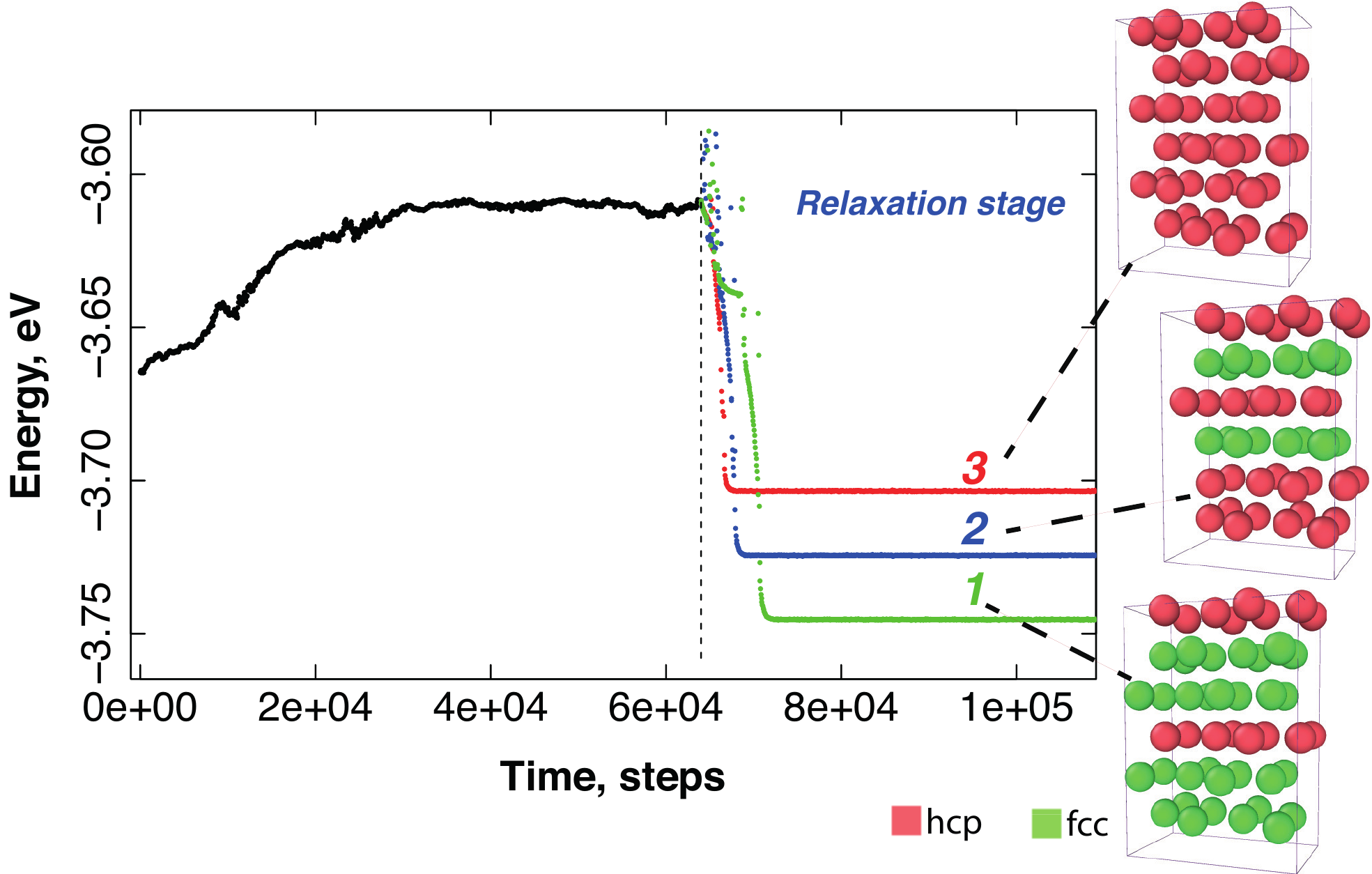}
  \caption{(Color online) The energy evolution of the reaction path from $hcp$ argon crystal to the layered structure of $fcc-hcp$.
  The final structures after relaxation of walkers:  1~-~the~layered structure of $fcc-hcp$ with a 66,67\% of $fcc$ layers, 2~-~the~layered structure of $fcc-hcp$ with a 33,33\% of $fcc$ layers, 3~-~the~original $hcp$ argon crystal.
  The red and green colored atoms are recognized as a $hcp$ and $fcc$ structures respectively.}
  \label{fig:energy_argon}
 \end{center}
\end{figure}

\begin{figure}[!t]
 \begin{center}
 \includegraphics[scale=0.3]{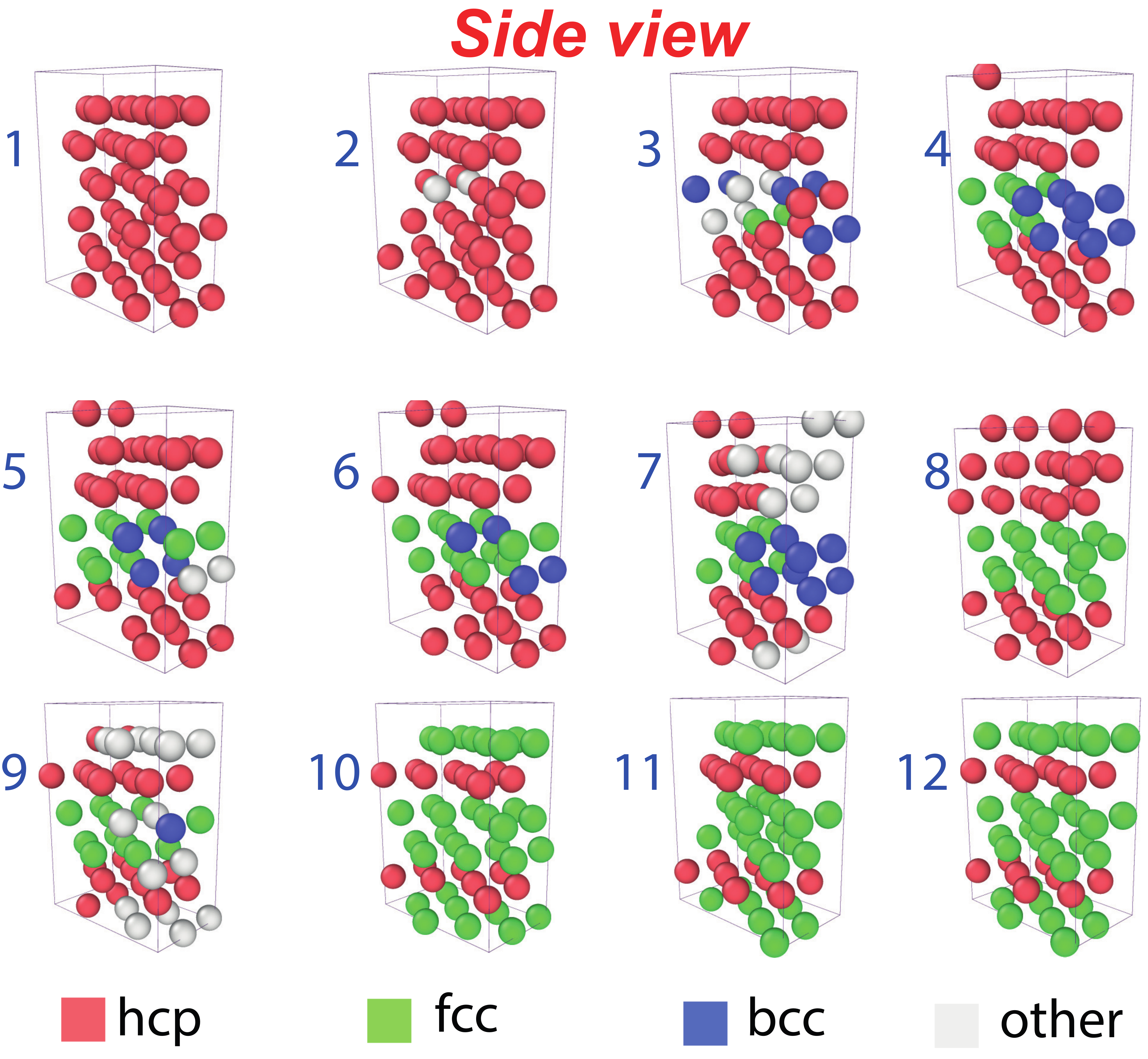}
  \caption{(Color online) The side view of the transition snapshots from hcp to the layered structure.}
  \label{fig:paths_argon}
 \end{center}
\end{figure}

\begin{figure}[!t]
 \begin{center}
 \includegraphics[scale=0.3]{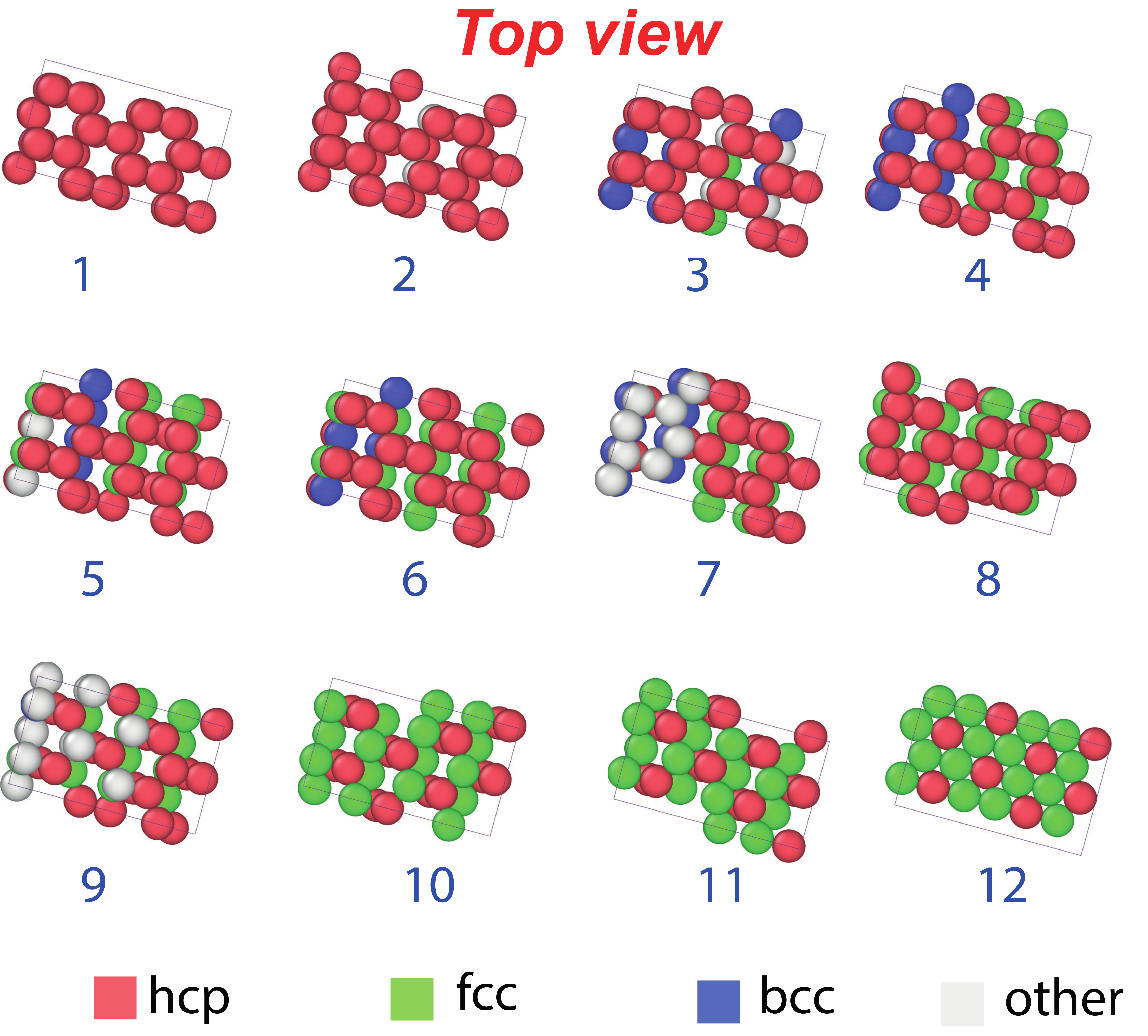}
  \caption{(Color online) The top view of the transition snapshots from hcp to the layered structure.}
  \label{fig:paths_argon_top}
 \end{center}
\end{figure}

The most interesting case is a simulation of solid-solid phase transition (PT) because this type of PT has a high energy barrier or/and it occurs under some special conditions \cite{Iron_bcc_hcp_fcc}. We applied AtomREM to the PT in the argon crystal. The theory for argon crystal \cite{argon_theory_1957} gives the two most stable structure for solid argon - $fcc$ and $hcp$, the authors of recent paper \cite{DFT_argon_pressure} performed the density-functional studies of argon at high pressure ($20-100~GPa$)  and also have found that properties of $hcp$ structure are very close to the $fcc$. Fig.~\ref{fig:energy_argon} shows the energy evolution and Fig.~\ref{fig:paths_argon} and Fig.~\ref{fig:paths_argon_top} demonstrate the side and top views of the atomic configurations during the simulation. For simulation we used the LAMMPS Lennard-Jones potential with $\epsilon = 0.238~kcal/mol=0.0103207~eV$ \cite{fcc_hcp_1},  $hcp$ cell with $2\times2\times3$ size with 48 atoms. To find the initial configuration, we used the Langevin mechanics simulation in the mode "Initialization" under {\it temp}$=0.5~K$ for 4000 time steps ($\tau = 0.0005$) and with 1600 walkers. To determine the initial configuration for the later ``Reaction" process, we compared the relaxation times among a group of walkers which exhibited two largest energy deviations for certain two neighboring atoms and selected the slowest among them (actually in the mode ``Reaction" the path was initiated by those atoms, as indicated by white color in the second snapshot of Fig.~\ref{fig:paths_argon}). The simulations of the reaction pathways were performed at the mode "Reaction" under the temperature $T=10^{-3}K$, $\tau = 5 \times 10^{-4}s$, 1728 walkers. The shape of the supercell was fixed during the simulation. To analyze atomic configurations and recognize the $hcp$, $fcc$ and $bcc$ structures, we used the OVITO tool (Open Visualization Tool \cite{Stukowski_2009,OVITO}) and adaptive common neighbor analysis with a variable cutoff \cite{CNA}. 

% The finding of the initial configuration was related to the 31st and 22nd atoms, which are neighbors and in the middle of the simulation box.The reaction path started from these atoms (colored white at the snapshot 2 in Fig.~\ref{fig:paths_argon} and \ref{fig:paths_argon_top}). The simulations of the reaction pathways were performed at the mode "Reaction" under the temperature $T=10^{-3}K$, $\tau = 5 \times 10^{-4}s$, 1728 walkers. The shape of the supercell was fixed during the simulation.% and with $NVT$ canonical ensemble.

%\begin{figure}[H]
% \begin{center}
 %\includegraphics[scale=0.3]{Figures/argon/structures.eps}
 % \caption{(Color online) The final structures after relaxation of walkers:  1-  the layered structure of fcc-hcp with a 66,67\% of fcc layers, 2 - the layered structure of fcc-hcp with a 33,3\% of fcc layers, 3 - the original hcp argon crystal.}
 % \label{fig:structures_argon}
% \end{center}
% \end{figure}

We have found the saddle point related to the original $hcp$ structure and the layered $fcc-hcp$ structures (Fig.~\ref{fig:energy_argon}). To make the walkers overcome the saddle point efficiently, {\it tempfin} was set relatively high, due to which the plot of Fig.~\ref{fig:energy_argon} showed fluctuation at the time step where $V({\bm x})$ was switched off. The energy barrier was $\sim 0.10/48=0.002083~eV/atom$ or $0.0480351089~kcal/mol$ for $hcp-fcc$ layered transition. This is reasonably smaller than the experimentally observed heat of fusion~\cite{argon_Flubacher_1961} of $0.2845~kcal/mol$.%, so our measure of energy barrier is less than $\Delta H_{fusion}$ at least three times.%, for the ejection point $E_b\approx0.1~kcal/mol$ (Fig.~\ref{fig:energy_argon}). 

A notable thing is that we realized the elementary processes of the solid-solid PT with large energy barriers without melting of the system. In the previous paper investigating the solid-solid PT in iron~\cite{Iron_bcc_hcp_fcc} the PT was induced by molecular dynamics with anisotropic compression so that the barrier height is lowered. AtomREM does not require such modification of the potential surface for realizing the elementary processes and enables us more straightforward comparisons to the experiments in arbitrary compression and strain setups.

A closely related method of the solid-solid PT has recently been demonstrated with the metadynamics~\cite{Laio01102002}, for example, in papers Refs.~\cite{PhysRevLett.119.015701, Meta_dynamics_nucl}. There, the nucleation problems have been investigated using an entropy-related order parameter and enthalpy as the collective variables. In such an approach the information of detailed atomic positions in the reaction paths are coarse-grained and there remains the possibility that the simulation results crucially depend on the choice of collective variables. Our method, without loss of the detailed atomic positions, could also be useful for seeking optimal collective variables.

\section{Conclusions and future work}
\label{sec:conclusion}
This article has introduced AtomREM, which has been developed in order to study the possible structural transformations to the unknown potential minima without introducing the empirical collective variables and artificial forces. The code package is released under the GNU General Public License 3 and available in a public software repository~\cite{AtomREM} including example cases. Various potential functions are available via the interface to LAMMPS molecular dynamics package~\cite{LAMMPS}. Demonstrations with the cyclobutane molecule and argon solid systems shows the performance of AtomREM. % and implements a highly-extensible FORTRAN 90/C++ based approach. AtomREM is parallelized using an MPI-based domain-decomposition approach and it is available as a public software repository \cite{AtomREM} that includes documentation and example cases. 

Currently, AtomREM cannot be combined with the density functional theory calculations for more accurate potential surface. To achieve the comparable accuracy, combination with the Deep potential molecular dynamics method~\cite{DPMD} is under way.

\bigskip
{\bf Acknowledgment} {\itshape We thank Taichi Kosugi for advice on the coding. This research was supported by MEXT as Exploratory Challenge on Post-K computer (Frontiers of Basic Science: Challenging the Limits). This research used computational resources of the K computer provided by the RIKEN Advanced Institute for Computational Science through the HPCI System Research project (Project ID:hp160257, hp170244, hp180184, hp190176).}

\newpage

\newpage

\newpage

\section{Refences}
\bibliographystyle{elsarticle-num}
\bibliography{CPC}

\end{document}